\documentclass[nonacm]{acmart} %

\usepackage{graphicx}
\usepackage{comment}
\usepackage{xspace}
\usepackage{color}
\usepackage{url}
\usepackage{hyperref}
\usepackage{cleveref}

\usepackage[T1]{fontenc} 
\usepackage{anyfontsize}

\newcommand\bigoh{{\mathcal O}\xspace}

\newcommand{\change}[1]{#1}
\newcommand{\taa}{\text{$\text{TA}3$}\xspace}
\newcommand{\sebi}{SeqBio\xspace}
\newcommand{\wca}{worst-case analysis\xspace}
\newcommand{\kmer}{$k$-mer\xspace}
\newcommand{\kmers}{$k$-mers\xspace}

\AtBeginDocument{%
  \providecommand\BibTeX{{%
    \normalfont B\kern-0.5em{\scshape i\kern-0.25em b}\kern-0.8em\TeX}}}

\setcopyright{acmcopyright}
\copyrightyear{2022}
\acmYear{2022}
\acmDOI{10.1145/1122445.1122456}

\begin{document}
\title{The theoretical analysis of sequencing bioinformatics algorithms and beyond}

\author{Paul Medvedev}
\affiliation{%
\institution{The Pennsylvania State University}
  \city{University Park}
  \country{USA}}
  \email{pzm11@psu.edu}

\begin{abstract} 
	The theoretical analysis of performance has been an important tool in the engineering of algorithms in many application domains. Its goals are to predict the empirical performance of an algorithm and to be a yardstick that drives the design of novel algorithms that perform well in practice. While these goals have been achieved in many instances, they have not been achieved ubiquitously across crucial application domains. 
I provide a case study in the area of sequencing bioinformatics, an inter-disciplinary field that uses algorithms to extract biological meaning from genome sequencing data. 
In particular, I give three concrete examples: two showing how theoretical analysis has failed to achieve its goals and 
one showing how it has been successful. 
I will then catalog some of the challenges of applying theoretical analysis to sequencing bioinformatics, argue why empirical analysis is not enough, and give a vision for improving the relevance of theoretical analysis to sequencing bioinformatics. By recognizing the problem, understanding its roots, and providing potential solutions, this work can hopefully be a crucial first step towards making theoretical analysis more relevant in sequencing bioinformatics and potentially other fast-paced application domains. 
\end{abstract}

\maketitle

\section{Introduction}

When I ask first year computer science undergraduate students how to quantify the speed of an algorithm, they look at me puzzled and tell me to just run the algorithm and see how long it takes. Such empirical analysis, in fact, is the most direct and natural way to measure algorithm performance. 
However, it has long been understood to have many shortcomings~\cite{Sedgewick2013}
and,
to overcome these shortcomings, computer scientists developed 
ways to theoretically analyze algorithm performance.
The most common technique for this is {\em traditional \wca}.
For example, we say that merge sort runs in $\bigoh(n\log n)$ worst-case time, which formally means
that there exists a constant $c$ such that for any large-enough input of $n$ elements, merge sort takes at most $cn\log n$ time.
Other more sophisticated techniques, 
such as parametrized analysis, average-case analysis, or semi-random models,
better capture the properties of real data~\cite{Roughgarden2019-dz}.
Additionally, theoretical analysis can be used to measure not just speed but other aspects of algorithm performance like memory usage or accuracy.  
When undergraduate students take an Algorithms course, they finally learn about the theoretical analysis of algorithms and
how to use it to capture general patterns of performance that empirical analysis does not.

The most direct impact of such theoretical analysis is in applied algorithms, i.e. algorithms which are implemented and applied to real data
(in contrast to complexity theory, where such analysis serves the purpose of understanding the hierarchy of problem, rather algorithm, complexity).
The goals of the theoretical analysis of applied algorithms, which we will denote by {\em \taa}, are twofold~\cite{Roughgarden2019-dz}.
One goal is to {\em predict} the empirical performance of an algorithm, either in an absolute sense or relative to others.
The second goal is to be a yardstick that drives the {\em design} of novel algorithms that perform well in practice.
\taa has achieved its goals with resounding success, being directly responsible for the design and performance prediction of many algorithms used in practice
(e.g. Dijkstra's shortest path algorithm).
Because of this success, Algorithms instructors typically jump into theoretical analysis with only a cursory justification of why it is needed.
To put it bluntly, the fact that \taa achieves the two goals has become a dogma of computer science.

However, fast-paced application domains pose a challenge to \taa.
In this paper, I use the field of sequencing bioinformatics (\sebi) as a case study, where I posit that \taa has failed to achieve its stated goals.
\sebi is an 
inter-disciplinary field that uses algorithms to extract biological meaning from sequencing data.
\sebi has revolutionized the life sciences, with algorithms developed by computer scientists 
(e.g. \cite{bowtie2,Bankevich2012-te})
enabling projects such as the  Earth Microbiome Project~\cite{gilbert2014earth},
the Vertebrate Genomes Project~\cite{vgp}, and the Cancer Genome Atlas~\cite{hutter2018cancer}.
It is also in the process of revolutionized healthcare, enabling projects such as Obama's Precision Medicine Initiative.
However, the vast majority of \sebi papers either do not perform \taa (e.g.~\cite{Bankevich2012-te}) or perform traditional \wca 
only to conclude that it is not a good predictor of the algorithm's performance in practice (e.g.~\cite{clever}).
In this paper, I will demonstrate two concrete examples of \taa's failure in \sebi:
the problems of genome assembly (\Cref{sec:assembly}) and structural variation detection (\Cref{sec:structural}).
I will also give one encouraging example of success: $k$-mer data structures (\Cref{sec:compact}).
I will then catalog some of the challenges of applying theoretical analysis in \sebi, argue why empirical analysis is not enough, and give a vision for improving the relevance of theoretical analysis to \sebi.
By recognizing the problem, understanding its roots, and providing potential solutions,
this work can hopefully be a crucial first step towards making \taa more relevant in \sebi and other fast-paced application domains. 

Before proceeding,
it is useful to make the distinction between direct and indirect influence of \taa in \sebi.
Without question, \taa can be credited with the development and analysis of methods which have
become part of \sebi's toolbox, e.g. integer linear programming, clustering algorithms, 
sketching techniques, machine learning, etc.
For the two problems we will investigate,
combinatorial algorithms for optimization problems with theoretical guarantees form the backbone of many 
tools (\cite{Bankevich2012-te,Simpson2010-br,opera,narzisi2014algorithmic} for assembly and~\cite{Medvedev2009-et} for structural variant detection).
However, in this paper we are concerned with {\em direct} influence,
i.e. situations where theoretical analysis was applied
to problems that are specific to \sebi, or at least for which \sebi was a major motivation.

\section{Accuracy of genome assemblers}\label{sec:assembly}
Sequencing Bioinformatics algorithms work with data generated by various instruments that, roughly speaking, repeatedly sample short substrings (called {\em reads}) 
from a long genome sequence~\cite{Mardis2017}\footnote{We note that in recent years, the reads have become much longer; however, many assembly algorithms predate this advance.}. 
The locations are chosen semi-uniformly at random but are not output by the instrument; the output only contains the string sequence of each read.
Moreover, the reads may contain errors and hence not be exact substrings of the genome.
Such a sequencing experiment can generate billions of reads at ever-decreasing costs.
There are many applications of sequencing technology but in this section we will focus on the genome assembly problem,
a classical \sebi problem.

\begin{figure}
\centering
\includegraphics[scale=0.8]{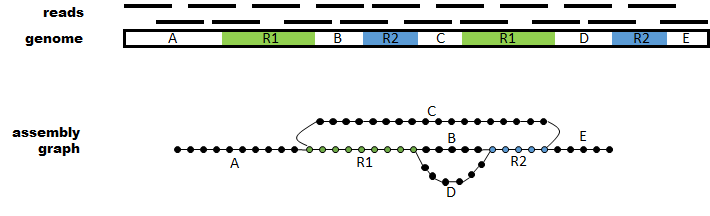}
\caption{
Cartoon illustration of a simple assembly algorithm.
The figure shows a hypothetical genome composed of several long segments. 
Two segments are repeated: the green segment R1 appears twice, as doe the blue segment R2.
The other segments are assumed to be repeat-free, in the sense that all the substrings of length above a certain threshold $k$ are unique.
A potential set of sampled reads is shown above the genome, lined up according to where they come from in the genome. 
Their location is not known by the algorithm but is shown here for clarity. 
A simple assembly algorithm would construct an assembly graph from the reads, shown at the bottom.
The nodes are all the $k$-mers (i.e. substrings of length $k$) appearing in the reads and there is an edge 
between a pair of \kmers that follow one another in at least one read.
The genome is a walk in this graph that covers all the vertices (A, R1, B, R2, C, R1, D, R2, E), however, there is more than one walk with this property (e.g. (A, R1, D, R2, C, R1, B, R2, E)).
A simple assembly algorithm would not risk making a mistake and would output only the walks in the graph that it is confident appear as sub-walks of any genome walk.
In this case, the output could be the spelling of the 7 walks A, B, C, D, E, R1, and R2.
}
\label{fig:assembly}
\end{figure}

The {\em genome assembly} computational problem is to take a sequencing experiment from a single genome and to 
reconstruct the full DNA sequence of that genome~\cite{Simpson2015}.
There are dozens of widely-used assemblers, with hundreds more at the prototype stage.
Genome assembly algorithms 
have enabled genome-wide studies of thousands of species
and have a biological impact that is hard to overstate.
The most important aspect of an assembler's performance is, arguably, its accuracy, 
since the resources required to collect a DNA sample  usually outweigh the computational resources of an assembler.
In this section, we will describe the various attempts to apply theoretical analysis to 1) predict the accuracy of assemblers
and 2) design more accurate assemblers.

\Cref{fig:assembly} gives a cartoon example of a simple assembly algorithm.
However, 
several practical factors complicate this simple picture. 
First, reads can have sequencing errors which introduce erroneous vertices and edges into the assembly graph.
Second, some parts of the genome are not covered by reads, introducing gaps in the graph.
Third, %
repetitive sequences are prevalent and make the graph structure more convoluted~\cite{nagarajan2009parametric,kingsford2010assembly}.
These and other factors make it challenging to keep the output of the assembler accurate.

One of the earliest theoretical measures of accuracy is the likelihood of the reads given an assembly.
The idea of building an assembler to maximize this likelihood was originally proposed in~\cite{myers1995toward} 
and later pursued in~\cite{medvedev2009maximum,varma2011improved,howison2014bayesian,bovza2015gaml}.
Much of the work centers on finding an appropriate likelihood function, i.e. 
one which models the intricacies of the sequencing process.
Unfortunately, these formulations have not directly led to any state-of-the-art assemblers,
leaving the design goal of \taa unfulfilled.
The reasons for this are not fully clear from the literature, but we will speculate in~\Cref{sec:challenges}.

I am not aware of any work that attempted to use likelihood to theoretically predict algorithm performance.
Some works did explore the idea of using a likelihood score to evaluate the accuracy of an 
assembly~\cite{vezzi2012reevaluating,rahman2013cgal,clark2013ale,hunt2013reapr,ghodsi2013novo,li2014evaluation}.
Though these tools have been widely used, they are not designed to give a theoretical accuracy of an assembler but,
rather, to be run on a concrete output.
As such, they cannot predict in advance how an assembler will empirically perform,
leaving the prediction goal of \taa unfulfilled.

A later approach~\cite{Bresler2013-yp} is to 
evaluate an assembler by the conditions under which it can fully reconstruct the original genome sequence.
Such conditions could, for example, be the number of reads needed or the highest error that could be tolerated.
This accuracy framework did lead to the design of a new assembler called Shannon that has been used in practice~\cite{Kannan_undated-we}, thus partially fulfilling the design goal.
However, these conditions rarely arise in practice
(i.e. the genome usually has too many repetitive sequences to be reconstructed completely and unambiguously). 
Therefore it is not clear if designing algorithms to optimize this would lead to other assemblers that perform well in practice. 
In terms of the prediction goal, this framework was also applied to theoretically predict the accuracy of some simple assembly strategies~\cite{Bresler2013-yp};
however, it has not been applied to predict the accuracy of any other assemblers used in practice.
The common challenge to all prediction attempts such as this one is that most assembly algorithms rely heavily on ad-hoc, hard-to-analyze heuristics.

One more recent approach is to measure what percentage of all the substrings that could possibly be inferred to exist in the genome are output by the assembler~\cite{Tomescu2017-me}. 
However, this framework has proven technically challenging to apply to real data,
e.g. to account for sequencing errors and gaps in coverage. 
It has not yet led to the design of a new assembler or to the accuracy prediction of assemblers,
though work is ongoing~\cite{cairo2020hydrostructure}.

In summary, theoretical analysis of assembler accuracy cannot be credited with the design of any of the widely used assemblers,
nor has it led to any theoretical analysis that can predict the accuracy of an assembler on real data.
In practice, assemblers are designed heuristically to perform well on a set of empirically observed metrics~\cite{narzisi2011comparing},
such as the lengths of the segments that the assembler reports or the recovery of genes whose sequences are conserved across different species~\cite{rhie2020merqury}.
Additional validation is performed by measuring the agreement with data from an orthogonal sequencing technology, 
where the sequencing errors have different patterns~\cite{rhie2021chasing}.
Assembly algorithms are usually developed to perform well on
publicly available datasets
and often validated by the exact same datasets.

The shortcomings of \taa have been felt in practice.
The assemblathon2 competition \cite{Bradnam2013-qa} performed an empirical evaluation of assemblers and found that the 
ranking of different assemblers according to their relative accuracy depended on the dataset, on the evaluation metrics being used, and on the parameter choices made in the evaluation scripts. These are exactly the shortcomings of empirical evaluation that \taa is intended to address. Moreover, the assemblers themselves are simply not as good as they could be, or, as one of the reviewers concluded, ``on any reasonably challenging genome, and with a reasonable amount of sequencing, assemblers neither perform all that well nor do they perform consistently'' \cite{Titus_Brown_undated-yv}. 
In the last five years, the practical situation has to some extent improved due to newer sequencing technologies that generate higher quality data. 
Nevertheless, the experience of the assemblathon2 competition is instructive to appreciate the limitations of solely-empirical analysis in \sebi.

\section{Accuracy of structural variation detection algorithms}\label{sec:structural}
Once a genome is assembled for a species, it forms what is called a reference genome.
Follow up studies then sequence different individuals of the same species
but do not perform a {\em de novo} genome assembly.
Instead, they catalog the variations between the sequenced genome and the reference, under
the assumption that the genome is unchanged in places where there is no alternative evidence. 
Variants that affect large regions of more than 500 nucleotides, called {\em structural variants}, are responsible for much genomic diversity and are linked to numerous human diseases, including cancer and a myriad of neurodevelopmental disorders~\cite{weischenfeldt2013phenotypic}.
Algorithms for detecting structural variation started to appear around 2008 (see \cite{Medvedev2009-et,Mahmoud2019-rw} for surveys) and a recent assessment identified at least 69 usable tools \cite{Kosugi2019-tu}.
As with genome assembly, the most important aspect of algorithm performance is, arguably, accuracy.

\begin{figure}
\centering
\includegraphics[scale=0.6]{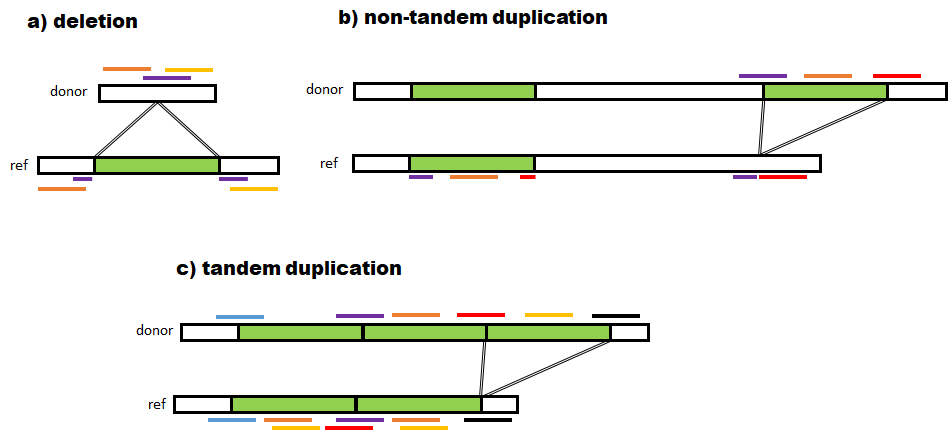}
\caption{
	Cartoon illustration of structural variation detection.
Each panel shows the sequenced donor genome (top rectangle) and the species reference genome (bottom rectangle). 
It shows the reads above the donor genome according to their origin position; 
it shows the reads below the reference according to where a string pattern matching algorithm would place them 
(in more technical terms, an alignment). 
In panel (a), the donor genome has the green region deleted; 
in panel (b), the donor has the green region duplicated and inserted far away in the genome;
in panel (c), the reference has two copies of the green region while the donor has three.
Observe how each event leaves behind a signature that can be detected by an algorithm.
In (a), the purple read is split into two partial matches, the position of each indicating the boundary of the deletion.
In (b), the purple and red reads each have two partial placements, with one end at the insertion location and the other at the edge of the duplicated sequence. 
In panel (c), all the reads have normal-looking placements; however, there are more reads mapping to the green region then one would expect if there was no duplication. 
}
\label{fig:sv}
\end{figure}

\Cref{fig:sv} illustrates the types of signatures that algorithms can use to detect structural variants.
What complicates the algorithm's task is that the same signature can sometimes be explained by alternate events, the signatures of multiple events can overlap, and
repetitive sequences can make it hard to find the correct location of a read.

The vast majority of tools are heuristics with no theoretical analysis of their accuracy.
There are some exceptions, when a probabilistic formulation is used to achieve a desired false discovery rate (e.g.~\cite{clever}).  In such cases, \taa can take some credit for the design of the algorithm. 
However, accuracy greatly depends on the type of variant (e.g. deletions are easier to detect than duplications) and on the location of the variant (e.g. repetitive sequences make variants harder to detect).
Thus, a useful analysis of accuracy requires a statistical model for the distribution of variant type, location, and relative frequency.
But coming up with realistic models is challenging, as
our understanding of the biological process that generates structural variants is limited.
Therefore, even when accuracy is predicted theoretically,
it does not correspond to what is observed in practice
because the models are too idealized~\cite{clever}. 
Thus, even in the limited cases where \taa has been applied, it has not achieved its prediction goals.

As with genome assembly, the algorithms used in practice suffer from many of the limitations that \taa is intended to address. 
Algorithms are typically evaluated empirically, using simulated data or an established benchmark.
Two recent studies assessing the empirical accuracy of algorithms~\cite{Kosugi2019-tu,Cameron2019-fj}
found that the tools suffered from low recall and the ranking of the tools according to accuracy varied greatly across different subtypes of variants. %
Empirical evaluation is hampered by the same lack of models that hampers theoretical evaluation and 
\cite{Cameron2019-fj} %
warned developers against considering ``simulation results representative of real-world performance.'' 
In fact, accuracy on simulated data is typically much higher for most tools then on real data \cite{Kosugi2019-tu}. 
As with genome assembly,
the problems described in \cite{Cameron2019-fj} are the types inherent to empirical-only evaluation: ``But with newly published callers [algorithms] invariably reporting favourable performance, it is difficult to discern whether the results of these studies are representative of robust improvements or due to the choice of validation data, the other callers selected for comparison, or over-optimisation to specific benchmarks.''

\section{Memory usage of \texorpdfstring{\kmer}{k-mer} data structures}\label{sec:compact}
Sequencing data is often reduced to a collection of $k$-long strings (called \kmers)
that are stored in a variety of data structures.
For example, breaking reads into constituent \kmers is part of many assembly algorithms (see~\Cref{fig:assembly}).
The exact data structure depends on the types of queries that need to be supported, the type of associated data that is maintained, and the source of the \kmer set.
Examples of queries include simple membership queries (i.e. is a \kmer present in the data structure?) and 
group membership queries (i.e. does a given bag of \kmers have at least $70\%$ of its \kmers present in the data structure?).
Examples of associated data include count information (e.g. how often does a \kmer occur in a set of reads?) 
or experiment information (e.g. given multiple experiments, which experiments contain the \kmer?).
Data structures to store \kmers have become ubiquitous in \sebi and form the backbone of hundreds of tools (for surveys, see~\cite{chikhi2021data,marchet2021data}).
The theoretical analysis of their memory is a 
rare bright light in the theoretical analysis of \sebi algorithms
and I discuss it here to illustrate \taa's potential for success in \sebi.

Many of the techniques used to analyse the memory used by \kmer data structures have been borrowed from the field of compact data structures~\cite{navarro2016compact}. 
The analysis of compact data structure memory differs from traditional worst-case analysis 
in that the higher order terms are often written without asymptotic notation
(e.g. $4n + o(n)$ instead of $\bigoh(n)$).
This helps distinguish algorithms whose memory differs by a constant factor, 
at the expense of a more technically involved analysis.

Using this type of analysis as a yardstick has led to the design of several \kmer data structures that perform well on real data and are included in broadly used software~\cite{boss,rainbowfish,minia,pufferfish,cqf,siren2017indexing}.
In one example, such an analysis led to the design of a compact representation of a popular \kmer data structure called the de Bruijn graph that uses $4n+o(n)$ bits, where $n$ is the number of edges~\cite{boss}.
This data structure uses very little memory in practice~\cite{spss} and forms the core of 
the widely used MEGAHIT assembler~\cite{megahit}. %
Another successful example of a \kmer data structure is the pufferfish index~\cite{pufferfish},
which %
forms part of the popular Salmon~\cite{salmon} software.
In addition to satisfying the design goal, this type of analysis has been used to correctly predict how much memory \kmer data structures will use in practice  and compare the relative performance of different methods (e.g. quantifying the tradeoffs between memory and query time of various indices).

Thus, theoretically analyzing memory by retaining the constant for the higher order terms has satisfied both the design and prediction goals.
Such a \taa success, though rare in \sebi, demonstrates that it is indeed possible for \taa to have an impact in \sebi.

\section{Empirical analysis is not enough}\label{sec:empirical}
In spite of the limitations of theoretical analysis, sequencing bioinformatics researchers continuously develop successful tools that have an enormous biological impact. 
A popular approach that fits both the accelerated development timeline and sidesteps the need for \taa 
is to reduce the problem to one from a toolbox of known approaches. 
A bioinformatician's toolbox includes black-box solvers for problems like clustering or integer linear programs; it also includes techniques like greedy algorithms or dynamic programming. 
This toolbox then forms the basis of incrementally designed heuristic algorithms.
These are rule based heuristics where the rules are incrementally improved by looking at where simpler heuristics fails on empirical data.
This is in contrast to a design driven by a mathematical understanding of the abstract problem with respect to a theoretical yardstick.

Empirical evaluation has its shortcomings but can nevertheless be very beneficial when done right. 
In particular, benchmark datasets and/or community competitions have been very useful. The Genome In A Bottle consortium for example has released both a benchmark sequencing dataset and a benchmark validation dataset for the problem of structural variant detection. Similarly, competitive assessment contests, such as 
assemblathon2~\cite{Bradnam2013-qa}, 
are designed to test submitted algorithms on strategically designed datasets. 
Such benchmarks and competitions aim to achieve similar design and prediction goals as \taa.
\change{Empirical analysis has in fact been the major driver of algorithm design in all the three examples covered here.}

Nevertheless, we have seen in~\Cref{sec:assembly,sec:structural} how \sebi suffers from serious problems that are hard to resolve with empirical analysis.
In many cases, benchmarks and competitions have not been developed, or have been developed years later than many of the algorithms (e.g. in structural variation).
In these cases, there is a proliferation of algorithms that perform well in the empirical validation of the authors but 
not in an independent evaluation. 
Note that the intent of the authors is not usually malicious; it is simply that the empirical validation approaches at their disposal are limited due to the lack of benchmarks. 
Even after the appearance of good benchmarks and competitions, algorithms that do well on these do not necessarily generalize well to other datasets; in fact, the incentives sometimes favor algorithms that overfit the data. Some assemblers, for example, are known to perform better on Human or {\em E.coli}, which are exactly the benchmarks commonly used for design and evaluation. 
On the other hand, algorithms that are designed to do well on a good (i.e. matching empirical observation) theoretical yardstick have the potential to be more generalizable, and the theoretical yardstick has the potential to predict algorithm performance on different datasets in a way that an empirical benchmark cannot. 
Thus, \taa can overcome the problems associated with relying solely on empirical analysis.

\section{The challenges of theoretical analysis in sequencing bioinformatics}\label{sec:challenges}

\change{
	Based on the preceding three examples, we can speculate on what factors have made theoretical analysis more challenging for the accuracy of genome assemblers and structural variation detectors than for the memory of \kmer data structures.
First, computer science is historically more applied to predicting memory rather than accuracy, which is more in the domain of statistics. 
Second, the two unsuccessful examples are closer to real data than the successful one, i.e. they are more subject to the whims of poorly understood biological processes.
Finally, it could simply be that compact data structures was a well developed field before its application to \sebi and bioinformaticians exploited that.
}

\change{Moving away from the concrete examples of this paper,} what is it about an application domain that makes the direct application of \taa so challenging? 
I can identify at least five challenges that are characteristic of \sebi but 
are general enough to possibly be present in other fast-paced application domains.
First, traditional worst-case analysis is in most cases too pessimistic when it comes to real data and fails to separate high performing algorithms, which take advantage of the structure of real data, from poorly performing ones that do not.  
This is in fact what many empirically successful algorithms do, as they stem from a deep understanding of the data followed by heuristics to exploit its structure. 
Such heuristics are difficult to analyze and are unlikely to be invented when the yardstick is traditional \wca.

Second, because applied researchers require a broad inter-disciplinary skill set, they often lack the technical expertise necessary to apply more sophisticated \taa techniques.
These techniques, sometimes called {\em beyond worst case analysis},
do in fact capture some of the complexities of real data~\cite{Roughgarden2019-dz}.
A good example of such a technique is smoothed analysis or, more generally, semi-random models, 
which make the analysis more realistic by assuming there is random noise forced upon any worst case instance.
However, these advanced techniques are rarely taught as part of the core CS curricula,
and applied researchers are typically exposed to theory only through introductory courses.
This affects the technical complexity of the \taa that they can perform.
For example, it is a rare case that a \sebi researcher can do a smoothed analysis of an algorithm. 
Being able to come up with a novel analysis technique is even more rare.

The third reason is that dataset sizes are growing at a rate faster than Moore's law~\cite{katz2022sequence}.
The usual justification for ignoring constants in traditional \wca is that a constant factor improvement in time or memory utilization will quickly become obsolete, as computing capacity grows exponentially.
But when the size of the data is growing faster than the computing capacity, a constant factor speedup may in fact be relevant for a long time.
This helps explain the success of the theoretical analysis of \kmer data structures, where the constant is kept 
(\Cref{sec:compact}).

The fourth reason is that algorithms in a fast-paced application domain are usually developed, analyzed, and applied under significant time pressure. 
In many cases, the algorithmic problem that a researcher is tasked with solving is only a small part of a more complex project in the application domain (e.g. biology).
Because the data and its underlying technology is rapidly evolving and changing, time is of the essence.
The researcher must work under time pressure to deliver a method that would analyze the data at hand and cannot afford to dedicate months of time to theoretically analyze an algorithm.
There are some notable exceptions of \sebi subareas where the data is stable enough so that the analysis and development of new algorithms has had enough time for complex \taa techniques to emerge (e.g. the edit distance problem),
but these cases are rare.

The fifth reason is that there is often an incentive to publish in venues belong to the application domain rather than in CS venues. 
In \sebi, for example, a paper will typically have more visibility if published in a biology journal rather than a CS bioinformatics conference. 
Domain scientists, however, rarely appreciate the difficulties of \taa, even if they lead to empirical breakthroughs. 
This especially affects early-career practitioners, who are incentivized to maximize visibility.

Because of these challenges, a \taa technique that is useful has to not only be predictive of empirical performance but also be easy to apply, easy to explain, and easy to understand. 
Thus, the theoretical analysis of algorithms in fast-paced application domains not only favors but in fact requires simplicity of the analysis technique. 
A simpler technique will be more trusted by domain scientists (e.g. biologists), more broadly understood and applied by practitioners, more easily taught to students, and more likely included in training curricula.
The challenge is to have a simple technique which nevertheless accurately captures empirical performance and is an effective yardstick for the development of empirically better algorithms.

\section{A vision for the future}

The first step to tackling the challenges of \taa in sequencing bioinformatics is to recognize the Theoretical Analysis of Applied Algorithms
 as its own research area, distinct from the design of the algorithms themselves. 
In \sebi journals or conferences, it is currently seen as a side note of algorithm development. 
Even in more theoretical \sebi venues, 
the value of a \taa contribution is not always appreciated. 
While a breakthrough result will likely be appreciated, a paper describing incremental progress on an algorithm is much more likely to be seen favorably than a paper describing substantial progress on an analysis technique. Moreover, there is generally an expectation that a \sebi paper, even a theoretical one, delivers a novel algorithm. In most cases, this is a valid expectation, but it is not always appropriate for \taa papers. 
Such publication challenges limit the formulation of \taa subproblems and are in general detrimental to progress in the \taa field. 

	The case study of \sebi can offer insights into other fast-paced application domains. 
	The challenges highlighted in~\Cref{sec:challenges,sec:empirical} can serve as a basis for reflection in those domains and it would be interesting to compare the \sebi experience with others.
\change{For example, the \taa techniques needed to tackle the challenges of structural variation detection may be very different from those needed to tackle the challenges of assembly; however,
		by aggregating these challenges across multiple domains, 
		we may find shared roadblocks and solutions. 
	}
	Recognizing \taa as a broad research area will help with the formation of a community of like-minded researchers and all the synergies that go along with it.
	Unfortunately, the current fragmentation of \taa research across multiple domains has been a bottleneck to progress.

The first step of a \taa research program could be to survey the literature for successful applications of \taa techniques.
This paper has surveyed the techniques in three sub-areas of \sebi, but a more thorough investigation is likely to turn up some more successful applications even within \sebi.
It will be useful to also identify successful \taa techniques from areas of bioinformatics that are not 
based solely on sequencing data, such as
whole-genome analysis and phylogeny reconstruction.
The toolbox of successful \taa techniques can then become the starting point of further research;
moreover, it can be added to the bioinformatics curriculum in computer science.

Viewing \taa as its own research field would allow researchers to 
focus on retrospective prediction of algorithm performance;
i.e. to develop \taa techniques using algorithms where the empirical performance is already known. 
For example, when the theoretical computer science community tackled the limits of the competitive ratio analysis technique for the online paging problem, the empirically best algorithm was already known; the challenge was to find the right technique to reach the same conclusion~\cite{Roughgarden2019-dz}. 
Similarly, a distinct \taa research program would not be afraid to tackle the question of performance prediction for short-read assembly, even though the empirically best methods have already been established.

The ultimate goal would be to develop \taa techniques that are simple yet predictive of real-world performance.
Within \sebi, these techniques could propel it forward and enable it to respond more quickly and accurately to the rapid evolution of sequencing technology.
Without investment in \taa research, \sebi and other application domains will continue to be hampered by the limitations of solely-empirical analysis of performance.

\paragraph{Acknowledgments:} 
This manuscript was inspired by the online videos for the 2014 class Beyond Worst-Case Analysis by Tim Roughgarden. 
I would like to thank Rob Patro, Jens Stoye, Alexandru Tomescu, and Fabio Vandin for reading an earlier version of this manuscript and providing great feedback, Michael Brudno for illuminating discussions on SVs, the twitter respondents at \href{https://twitter.com/pashadag/status/1279232810407124992}{https://twitter.com/pashadag/status/1279232810407124992}, and Jouni Siren for helping me understand the role of BOSS in VG.
I would like to also thank an anonymous reviewer for for suggesting publication incentives as one of the challenges in~\Cref{sec:challenges}.
This material is based upon work supported by the National Science Foundation under Grants No. 2138585, 1453527, and 1931531.
Research reported in this publication was supported by the National Institute Of General Medical Sciences of the National Institutes of Health under Award Number R01GM146462. 
\bibliographystyle{plain}
\bibliography{tasba}

\end{document}